# Wayne State University's Dan Zowada Memorial Observatory: Characterization and Pipeline of a 0.5 Meter Robotic Telescope

Robert Carr[1], David Cinabro[1], Edward Cackett[1], David Moutard[1], Russell Carroll[1]

[1] Department of Physics & Astronomy, Wayne State University, Detroit, United States of America

E-mail: Robert.S.Carr@wayne.edu

## Abstract

Wayne State University's Dan Zowada Memorial Observatory is a fully robotic 0.5m telescope and imaging system located under the dark skies of New Mexico.  The observatory is particularly suited to time domain astronomy: the observation of variable objects, such as tidal disruption events, supernovae, and active galactic nuclei.  We have developed a software suite for image reduction, alignment and stacking, and calculation of absolute photometry in the Sloan filters used at the telescope.  Our pipeline also performs image subtraction to enable photometry of objects embedded in bright backgrounds such as galaxies.  The 5 sigma detection limit of the Zowada Observatory for integration of 16 x 90 second exposures is 19.0 magnitude in g-band, 18.1 magnitude in r-band, 17.9 magnitude in i-band, and 16.6 magnitude in z-band.  For a 3 sigma detection limit, measurements may be performed with greater uncertainties as deep as 19.9, 19.1. 18.9 and 17.5 magnitude in griz bands, respectively.

## 1. Introduction

Wayne State University's Dan Zowada Memorial Observatory, hereafter Zowada, is located in the high desert of the southwest corner of New Mexico and close to the border of Arizona, just outside the settlement of Rodeo, 2019 population of 108 (U.S. Census Bureau 2021). Rodeo (31°50′13″N, 109°01′54″W) is located about 80 miles due east of Tombstone, Arizona and about 160 miles due east of Kitt Peak Observatory. Zowada's location provides excellent dark sky viewing and generally stable weather conditions.

Zowada is ideal for time domain astronomy.  We use Zowada to research tidal disruption events (TDEs) and follow up on transients identified by the Dark Energy Spectroscopic Instrument (DESI). We also perform reverberation mapping of active galactic nuclei (AGN).  Our advanced undergraduate astronomy majors employ Zowada in their Astronomical Techniques Laboratory.

This paper describes the methods, capabilities, and limitations of Zowada.  The paper is arranged as follows. Section 2 describes the system and optical path and Section 3 characterizes the observatory site and pointing accuracy. Section 4 discusses implementation of our image processing pipeline.  Section 5 provides analysis of performance of Zowada.  Finally, Section 6 discusses considerations for researchers employing the Zowada Observatory.

## 2. System and Optical Path

Zowada is a fully robotic observatory. The DC3 Dreams ACP web form-based interface is used for observatory scheduling. Users program the observatory for observations and receive image data over internet. Three onsite computers manage observatory operations.

Zowada is housed within a 12.5 foot motorized clamshell dome. We use a Davis Instruments Vantage Pro2 weather station to measure wind speed, temperature and pressure, and a Boltwood Cloud Sensor II to measure sky temperature to determine the level of cloud cover: these support weather safe operations and automated closure of the observatory dome to protect against rain and wind events. We use an SBIG All-Sky 340 camera to observe and document sky conditions every five minutes day and night. During a summer 2019 upgrade, we added a windbreak to reduce wind driven vibrations to allow for operations during higher winds and a backup diesel generator to reduce downtime and equipment risk during power interruptions.

The telescope is a 0.51m (20 inch) corrected Dall-Kirkham telescope manufactured by PlaneWave Instruments, of Adrian, Michigan. The telescope has a focal length of 3454mm and focal ratio of f/6.8. The telescope is mounted on a Paramount ME II managed by Software Bisque software. Focusing is driven by the Maxim DL software. The camera is a Finger Lakes Instruments ProLine 23042 BB with E2V CCD230-42 sensor. The back-illuminated CCD is 2048 x 2048 pixels, with a physical pixel size of 15 micrometers by 15 micrometers. The pixel scale is 0.9 arc second per pixel. The field of view is approximately 30 arc minutes x 30 arc minutes. The wavelength range of the camera is 2600 Å – 10,000 Å. Quantum efficiency is better than 85% from 4600 Å - 5600 Å and exceeds 70% from 4000 Å - 7400 Å; quantum efficiency drops rapidly outside those bounds.

A 10-position filter wheel is mounted between the telescope and the camera. The wheel includes Astrodon Gen 2 Sloan u' g' r' i' and Pan-STARRS $z_s$ filters (hereafter ugriz) to support scientific observations. The wavelength centers are at 3540 Å, 4770 Å, 6215 Å, 7545 Å and 8700 Å, respectively. Transmission efficiencies of the Sloan filters are 95% in u, 98% in g, and 99% or better in r, i and z wavelength ranges. The wheel also includes Hα, luminance, and RGB filters for astrophotography.

Observations at Zowada currently focus upon g, r, i and z filters as this is where the net efficiency of the optical path is the best. Imaging in u-filter is inefficient and difficult at magnitudes fainter than about 10. Zowada u-filter observations will generally not be discussed further.

## 3. Characterization

### 3.1 Pointing Accuracy

The telescope is mounted on a Paramount ME II managed by Software Bisque software, with coordinates entered by users via the DC3 Dreams ACP web form-based interface. Pointing accuracy is potentially important to ensuring that the appropriate target is imaged and identified for photometry.

Pointing error is assessed by comparing input coordinates of the target object to the coordinates of the same object's centroid in the science image. Sources of pointing error include user coordinate rounding, mount errors and object proper motion.

RMS pointing error is 1.44 arc seconds in RA and 1.22 arc seconds in DEC. The pointing errors are consistent and biased in one direction: target coordinates in the image are generally greater than input coordinates in RA and DEC, as shown in Figure 1. These pointing errors compare to the pixel scale of 0.9 arc second per pixel, corresponding to roughly 1-1/2 pixels. Pointing errors are well within the typical point spread distribution of target objects. Pointing errors do not affect operation of the photometry pipeline, which identifies image and catalog objects within a radius of 2.0 arc seconds, or 2.2 pixels.

We have not assessed mount tracking accuracy during exposures. We rarely employ exposures greater than 300 seconds and have had no noticeable tracking problems for exposures of 300 seconds or less.

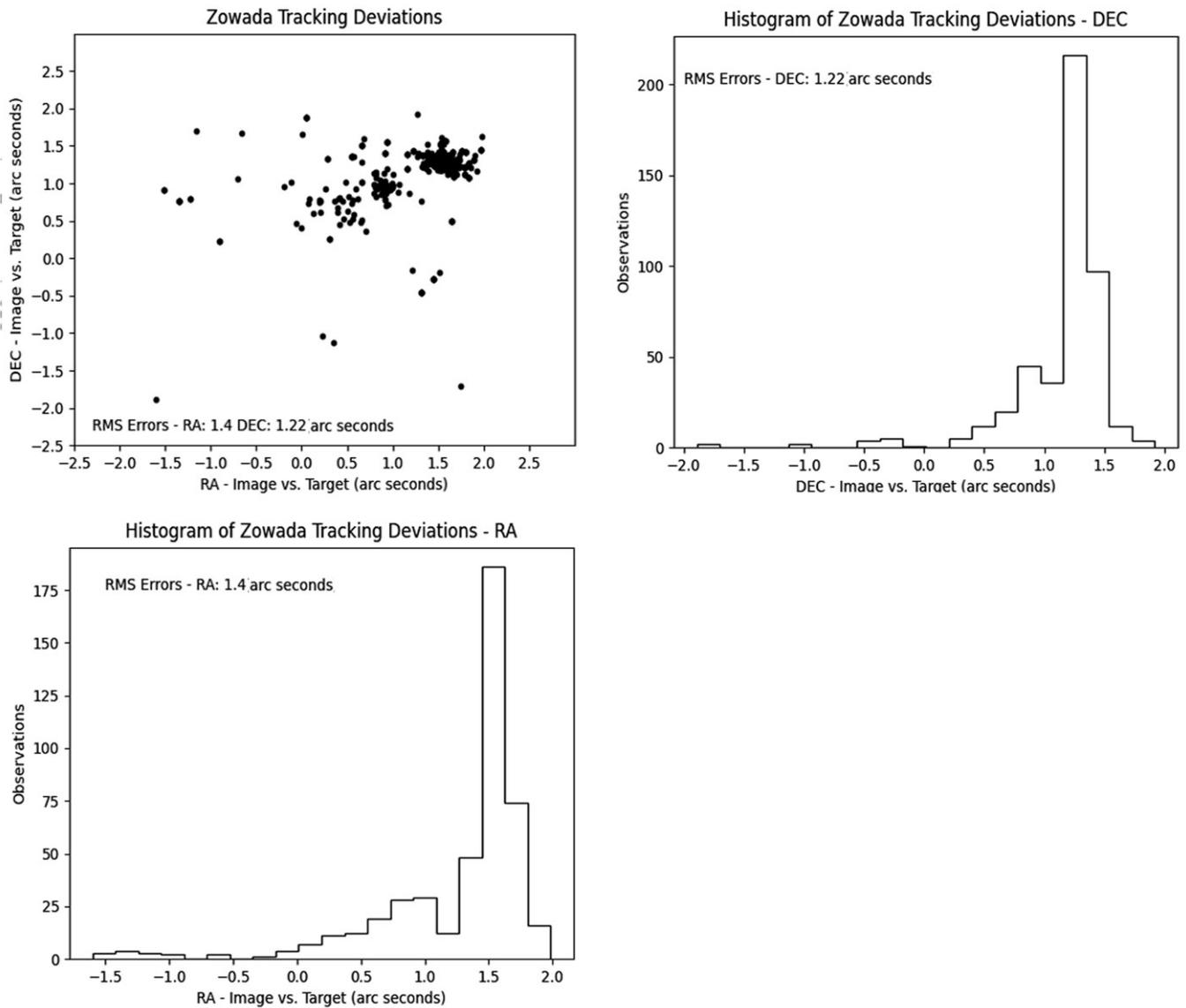

**Figure 1.**  Scattergram of Zowada pointing errors and histograms of errors in RA and DEC.  RMS tracking errors of 1.4 arc seconds in RA and 1.22 arc seconds in DEC compare to the pixel scale of 0.9 arc second per pixel of the camera.  The pointing error is well within the point spread of target objects.

*3.2 Observing Conditions*

Zowada's location in the high desert provides excellent dark sky viewing and generally stable weather conditions. We performed observations during 88% of the available nights excluding summer shutdown, when Zowada is closed for monsoon season.  The average night included 5.0 hours of observations.

To assess variability of seeing as measured by FWHM at Zowada, we reviewed observations over the period January 16, 2019-June 15, 2020, from installation of the Finger Lakes camera to the 2020 summer shutdown.  Observations were sampled at 30 minute intervals to avoid over-sampling of short and repeated exposures.  Over the 18 month period, FWHM ranged from 1.2 to 9.0 arc seconds.

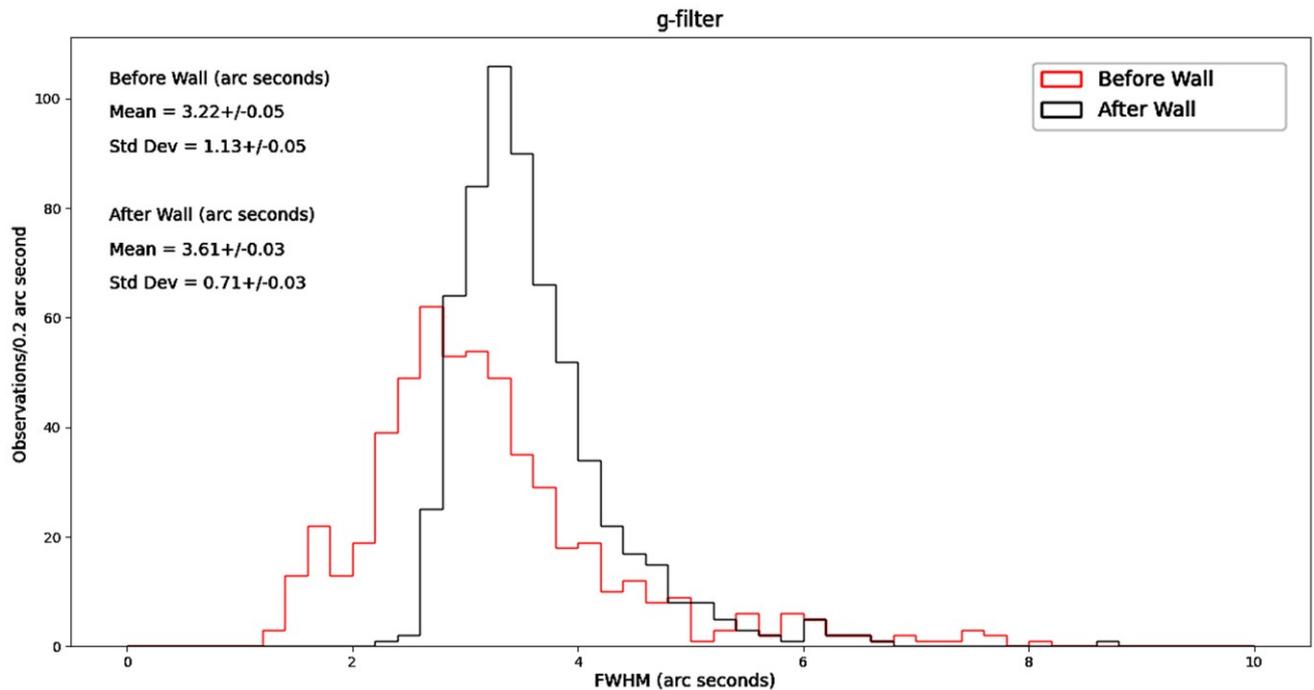

**Figure 2.** Effect of the wall around the Zowada Observatory.  After construction of the wall, mean FWHM increased while the range of FWHM experienced decreased.  FWHM is compared for January 16 - June 15, 2019 and January 16 - June 15, 2020.

During summer shutdown in 2019, we installed a wall around the Zowada Observatory to reduce the number of clear nights lost to gusts by increasing the threshold wind gust speed considered safe for viewing.  We evaluate the effect of the wall upon FWHM and observation time by comparing the period before construction of the wall in 2019, from January 16 to June 15, 2019, to the same period after construction of the wall in 2020.  Figure 2 shows the mean FWHM increased from 3.22 to 3.61 arc seconds in g-band, while the standard deviation of the FWHM distribution decreased from 1.13 to 0.71 arc seconds.  The wall has reduced observations at both low and high FWHM; Zowada now experiences no observations with FWHM less than 2 arc seconds and fewer observations with FWHM worse than 6 arc seconds.  The increase in mean FWHM since installation of the wall may be due to increased turbulence above the dome due to airflow over the wall.

The wall accomplished its purpose of increasing observatory time.  There were about 11% more observations in 2020 compared to the same period before installation of the wall in 2019.  Increased observation count reflected increased operating schedules due to improved weather protection.  Different weather conditions or observing schedules might also have contributed.

Our measurements for the remainder of this article evaluate FWHM since the installation of the wall.  Our measurements of mean FWHM are 3.61, 3.46, 3.31 and 3.14 arc seconds in g, r, i, and z-bands, respectively.  Seeing appears to weakly improve with wavelength as expected from theory (Coulman 1985).  FWHM can show significant variability within a week or even a day.  We have not found minimum/"best" FWHM to be related to ambient temperature or to occur at a consistent time of night.

At Zowada's location, the sky background level at high elevation in new moon is about 17.9 magnitude per arc second in g-band, 16.4 magnitude per arc second in r band, 15.4 magnitude per arc second in i band and 15.1 magnitude per arc second in z band.

## 4. Image Processing Pipeline

*4.1 Introduction*

The dome, telescope mount, filter wheel and camera at Zowada are fully robotic. Users enter objects for imaging through the DC3 ACP web-form interface. Objects are automatically scheduled for observation, prioritized based upon user entered priority weighting, monitoring interval, object altitude, and available observatory time. Exposures of acquired images typically range between 30 - 300 seconds. An internet link relays the acquired raw images to observatory users.

We developed a series of software modules in Python to perform image processing and analysis. The pipeline employs various modules from Astropy (Astropy Collaboration et al. 2013; Astropy Collaboration et al. 2018) and Photutils (Bradley et al. 2020) and ideas and structure from OKeeffe (2017).

We perform image reduction upon raw science frames in the usual way to produce bias and dark subtracted, flat-fielded images. Our image reduction uses the most recent available calibration frames: weather conditions permitting, calibration frames are taken in the morning at the conclusion of the nightly observing session which produced the science image. Upon completion image reduction, we align and stack the science frames to improve signal-to-noise ratio (SNR) of the final science frames.

Relative or absolute photometry are performed upon the resulting images. If we are studying an object located on a bright background, we first perform subtraction to isolate the object. Image subtraction removes the surrounding background to allow photometry on the object of interest. This is useful for study of objects such as TDEs within a galactic bulge

*4.2 Relative Photometry*

We perform relative photometry to determine the relative flux of a variable object and to construct light curves of objects that have been observed over several epochs. Relative photometry works by comparing the count rate of the variable object of interest with the count rates of non-variable comparison stars in the same field of view. We determine count rates using simple aperture photometry on bias and dark subtracted, flat-fielded images. The comparison stars are assumed not to vary from one epoch of observations to another, and so a scale-factor is calculated for each epoch so that the sum of the count rates from the comparison stars remains constant. The scale factors account for changes in the count rate due to changes in seeing between epochs. We apply these scale factors to the count rate of the target of interest to determine the target's light curve. To set the absolute flux scale, the magnitudes of the comparison stars can be obtained from appropriate catalogs.

The Zowada Observatory has been performing near daily monitoring of AGN in order to perform continuum reverberation mapping, where time lags are determined between the AGN light curves at different wavelengths (see Cackett, et al (2021) for a review of reverberation mapping). The AGN that have been monitored are relatively nearby and bright Seyfert 1s, which have B-band magnitudes typically in the 14 to 15 magnitude range. We usually obtain observations in all the ugriz bands in pairs of 300 second exposures in g, r and i bands and in three or four exposures in the less-sensitive u and z bands. With these exposures at these magnitudes, it is normally the case that the standard deviation in the comparison star light curves is the larger source of uncertainty over the statistical uncertainty in the count rates. This limits the photometric accuracy to around 1% in g, r, and i bands and 2-3% in u and z bands. The typical variability amplitude of AGN is several times this in u, g, r, and i bands. However, AGN are less variable at longer optical wavelengths, and in the z band the variability amplitude can sometimes approach the photometric uncertainty, making time lag measurements in that band more challenging.

A more detailed description of the application of relative photometry to the Zowada Observatory monitoring data of AGN, along with example light curves, is given in Cackett et al. (2020). Figure 3 provides a g-band light curve for AGN Mrk 142 adapted from this source. Other examples of Zowada Observatory data used for reverberation mapping are Kara et al. (2021) and Vincentelli et al. (2021).

*4.3 Absolute Photometry*

We perform absolute photometry upon a target object by comparing the target image's instrumental flux to a sample of reference objects. The photometry module calculates a linear relation of instrumental flux to magnitude based upon the reference objects. The magnitude of the target object is calculated based upon this equation and its instrumental flux.

We employ aperture photometry to obtain our instrumental fluxes. Circular apertures are used based upon FWHM of the science image. Target object flux is calculated based upon the flux in the circular aperture, after subtraction of the median background flux of a corresponding area of a surrounding annulus. The aperture radius is typically taken at 2.0 times the FWHM. At the mean FWHM in g-band of 3.61 arc seconds, mean aperture radius in r-band is 7.22 arc seconds, the inner radius of the annulus is 10.83 arc seconds and the outer radius is 14.44 arc seconds. We subtract median rather than mean instrumental flux measured in the background annulus because this has been found to reduce the effect upon magnitude measurements of faint objects lying within the background annulus. Alternative apertures and annuli were investigated but found to provide equivalent measurements.

To construct the line for calculation of the magnitude of the target object, the photometry module builds a sample of nearby (typically within 0.2 - 0.4 degrees) comparison objects of similar accepted measurements of luminosity (typically +/- 1.0 or 1.5 magnitudes) to the target object. Accepted magnitudes for comparison objects are stipulated to correspond to measurements by Zowada of instrumental flux for these objects. The photometry module filters the comparison objects for like objects, i.e. stars are compared to stars and galaxies to galaxies, and eliminates objects of insufficient photometric quality. We perform additional filtering of comparison objects to exclude objects not suitable for photometry; these "outliers" most often include objects with point spread functions (PSFs) that overlap within the 0.9 arc second per pixel scale of the Zowada camera. The photometry module excludes a comparison object if the magnitude that would be calculated for the object based upon its instrumental flux is more than three standard deviations greater than or less than its accepted catalog value. We have not calculated extinction or color transformation coefficients and are not planning to do so. Our frame by frame calibration method works better where we have sufficient number of calibration stars in the science image. For a given measurement, the sample size of nearby comparison reference objects ranges from about a dozen at 13 magnitude to over a hundred at 16 magnitude. Figure 4 illustrates the construction of the best fit line for calculation of magnitude of sn2020uyn; a sample of APASS accepted magnitudes and Zowada measurements of instrumental flux for 248 nearby stars was used to set the line..

Comparison reference objects are taken, when possible, from the Sloan Digital Sky Survey (SDSS) Catalog, Data Release 12 (Alam 2015). Our pipeline accesses this catalog through Vizier (Ochsenbein et al 2000; Ochsenbein 2015)\ using Astroquery (Ginsburg et al. 2019). SDSS is the database employed by our primary research partners and provides the ability to compare observations taken with Zowada's ugriz filter set against accepted catalog values over a wide range of magnitudes. The SDSS catalog includes objects fainter than roughly magnitude 9 over much of the Northern sky. Detailed information is available regarding object type and numerous specific photometric "flags", which provide detailed information on the quality and limitations of the underlying photometry. Our pipeline uses these flags to filter reference objects for "clean" photometry and exclude objects whose flags indicate saturation, blended objects, or various issues with the observed PSF. Filtering upon the photometry flags including saturation excludes objects brighter than 15 magnitude and about 20 - 35% of the remaining fainter objects, while significantly improving the standard deviation around measured magnitudes. Drawbacks of the SDSS catalog are that its sky coverage is limited and "clean" photometry is available only for objects fainter than 15 magnitude.

When we need to perform photometry in areas of the sky not covered by the SDSS catalog, we take reference objects from the American Association of Variable Sky Observers (AAVSO) Photometric All-Sky Survey (APASS), Data Release 10 (Henden et al. 2018). The APASS catalog includes objects from about 7 - 17 magnitude in the ugriz filter set. This catalog covers a complementary fraction of the sky and provides magnitudes for brighter objects than the SDSS catalog. The APASS catalog is published to provide standards for the measurement of variable objects and is constructed to only retain objects sufficiently constant in flux. However, it does not provide the detailed photometric data of the SDSS catalog. Between them, the SDSS and APASS catalogs cover much of the sky visible to the Zowada Observatory.

*4.4 Image Subtraction*

We include an image subtraction module in our pipeline to enable us to measure objects that are positioned against a luminous background. This image subtraction begins with a reduced Zowada image that contains a transient and removes the host galaxy, leaving only the transient. This allows us to measure absolute photometry and construct light curves for objects such as supernovae and TDEs, which may be otherwise obscured by their host galaxy.

The image subtraction module is performed upon science images generated from the reduction and stacking of raw files, as discussed above. The first step of the image subtraction module is to build a template of the field for subtraction. We download template images using the *panstamps* (Andika et al. 2020) software. This Python software gathers images of a requested field from the Pan-STARRS (Magnier et al. 2013) archive, in the requested filters. These template images are aligned to the science images from the Zowada Observatory using the *reproject* (Robitaille et al. 2020) Python module. This module rotates the template and resamples the Pan-STARRS image to be the same as the Zowada image. Once aligned, masks are made for both the Zowada image and the template. These masks account for hot and dead pixels, and edge artifacts that arise from shrinking and rotating the larger Pan-STARRS templates.

Before we can subtract the aligned template from the science image, we must first match the PSF and the scaling of the two images. This is done using the High Order Transform of PSF and Template Subtraction (*HOTPANTS*) (Becker 2015). We first measure the PSF of both images using the Astromatic package *PSFEx* (Bertin 2011). Then we provide *HOTPANTS* with some basic information to allow it to transform the template to match the science image. This information includes the masks and the size and number of the Gaussians of the PSFs used to smear the template image. The width of the Gaussian in the kernel is calculated as suggested by the *HOTPANTS* documentation by $\sigma_m = \sqrt{\sigma_t^2 + \sigma_s^2}$. Here $\sigma_m$ is the size of the central Gaussian used by *HOTPANTS*, and $\sigma_t$ and $\sigma_s$ represent the size of the PSF for the template and science images respectively, as calculated by *PSFEx*. If there is a transient that is above Zowada detection limits, then *HOTPANTS* will return a blank field with only one source remaining, as illustrated by SN2020uyn in Figure 5: this is the transient under study. There are generally some artifacts from subtraction, but there is rarely any ambiguity in detecting the transient itself. We can also use the Astromatic package *SExtractor* (Bertin 2011) to confirm the detection of a transient.

Once the subtracted image is generated, we then take a 160x160 pixel cutout around the source from the subtracted image. This cutout is inserted into the same region of the reduced and stacked Zowada image. This allows us to place the isolated source back in its original field to have a proper comparison for photometry. We perform absolute photometry upon the transient in the final image as described previously. Light curves are constructed from time series of the photometric data, with an example shown in Figure 6.

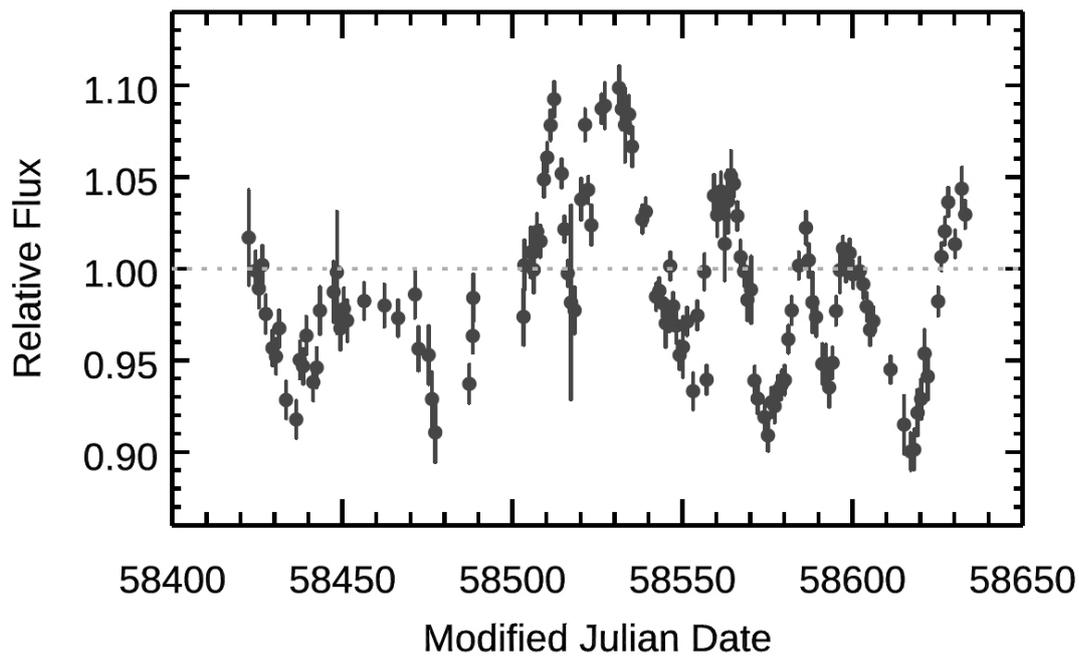

**Figure 3.** Light curve from relative photometry. The g-band light curve of AGN Mrk 142 was obtained using relative photometry (adapted from Cackett et al. 2020). Clear variability on timescales of days to weeks is easily detected.

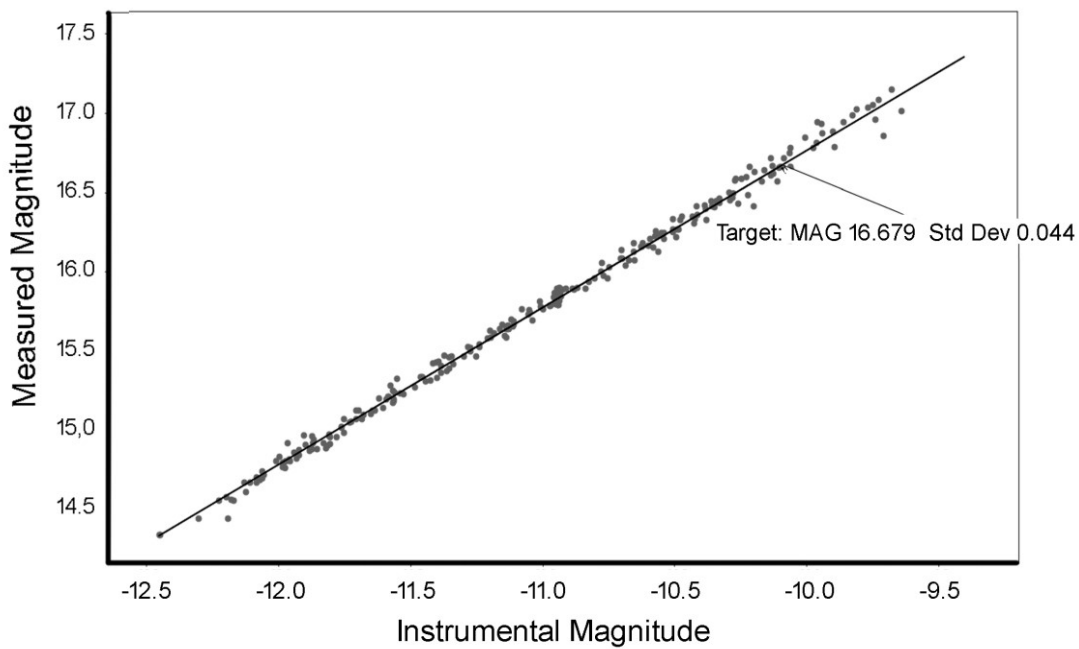

**Figure 4.** Best fit line for Zowada magnitude measurement for a particular target object. This line was constructed in g-band for absolute photometry of SN2020uyn. The line was fitted based upon Zowada instrumental flux and APASS magnitudes for 248 nearby stars within +/- 1.5 magnitudes of the target object's estimated luminosity..

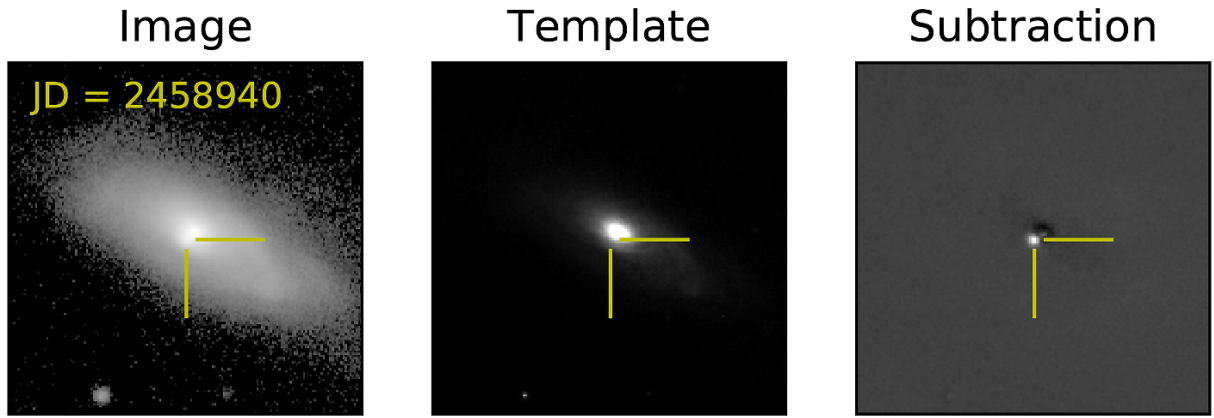

**Figure 5.** Image subtraction pipeline. The pipeline is performed on supernova SN2020uyn at JD 2459139. From left to right are the reduced Zowada Image, the Pan-STARRS template, and, finally, the image after subtraction, leaving behind a source. Artifacts from subtraction generally look distinct from real sources.

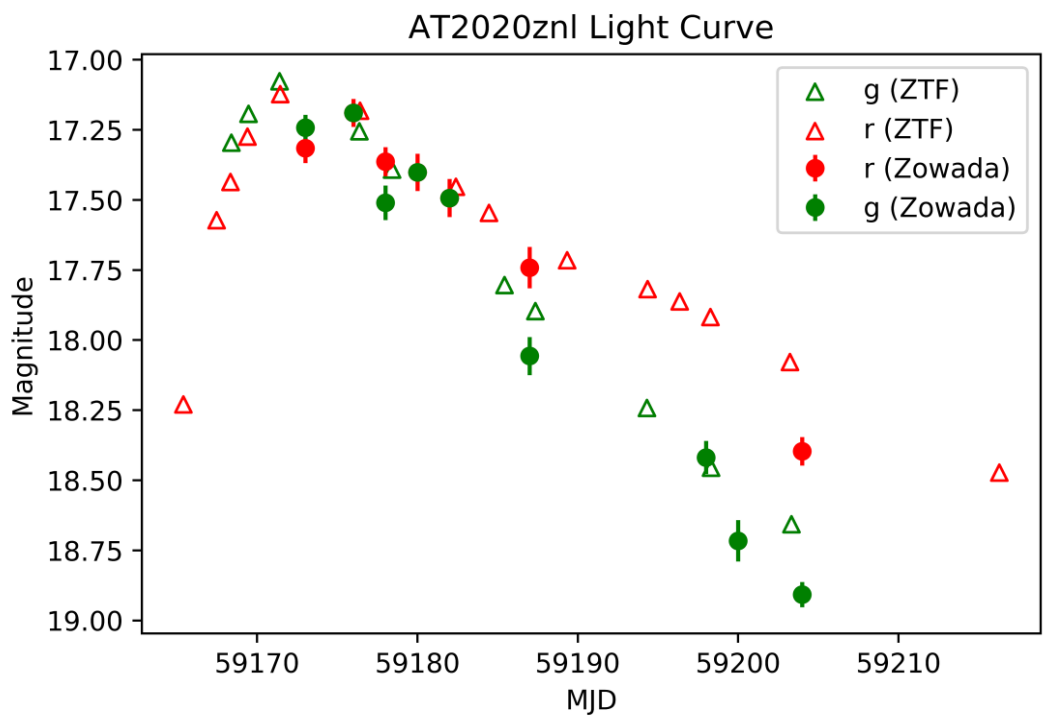

**Figure 6.** Light curve. Preliminary analysis assembled for supernova AT2020znl from a time series of Zowada photometric data. Solid dots represent Zowada data using a preliminary version of the method described above. Open triangles show data on the same object collected by the Zwicky Transient Facility (ZTF).

## 5. Data Analysis

*5.1 Magnitude Measurements*

We may perform standalone measurements of an object's luminosity or, more often, compile these measurements over time to build a light curve for an object of varying flux. Current research at the Zowada Observatory relies upon light curves, including observations of TDEs and AGN.

We assess Zowada's capability for measurement by comparison with SDSS catalog measurements and by its limiting magnitude. Accuracy of measurements is assessed by comparison of Zowada measurements to SDSS catalog measurements because the photometry module measures the magnitude of an object within a science image by comparing its luminosity with nearby objects drawn from the SDSS catalog.

Our magnitude measurements at the Zowada Observatory track closely to those in the SDSS catalog over a wide range of magnitudes and objects, as shown by measurement of magnitudes for 353 stars in g-band in Figure 7. Measurements performed with longer integration time of 16 x 90 second exposures (1440 seconds) have smaller errors compared to the SDSS catalog and a smaller spread of errors than do measurements performed with shorter integration times of generally 4 x 60 or 4 x 90 second exposures (240 - 360 seconds). Results are similar in r, i, and z bands. Errors for measurements of objects indicated by SDSS as having clean photometry were lower at low magnitudes than at high magnitudes, as expected. SDSS photometry at 13-14 magnitude is flagged as saturated and potentially unreliable; we believe this leads to higher magnitude measurements by SDSS than Zowada for these brighter objects.

RMS errors for Zowada measurements compared to SDSS catalog are less than 0.2 magnitude through 18.5 magnitude in g and r bands, 18.3 magnitude in i-band, and 16.5 magnitude in z-band. Measurement in z band is more challenging because of the lower net optical path efficiency. Errors tend to increase with increasing magnitudes and with i and z bands compared to r and g bands. As expected, Zowada is able to measure deeper magnitudes for longer integration times, while the distribution of errors widens for higher magnitudes and narrow for longer integration times.

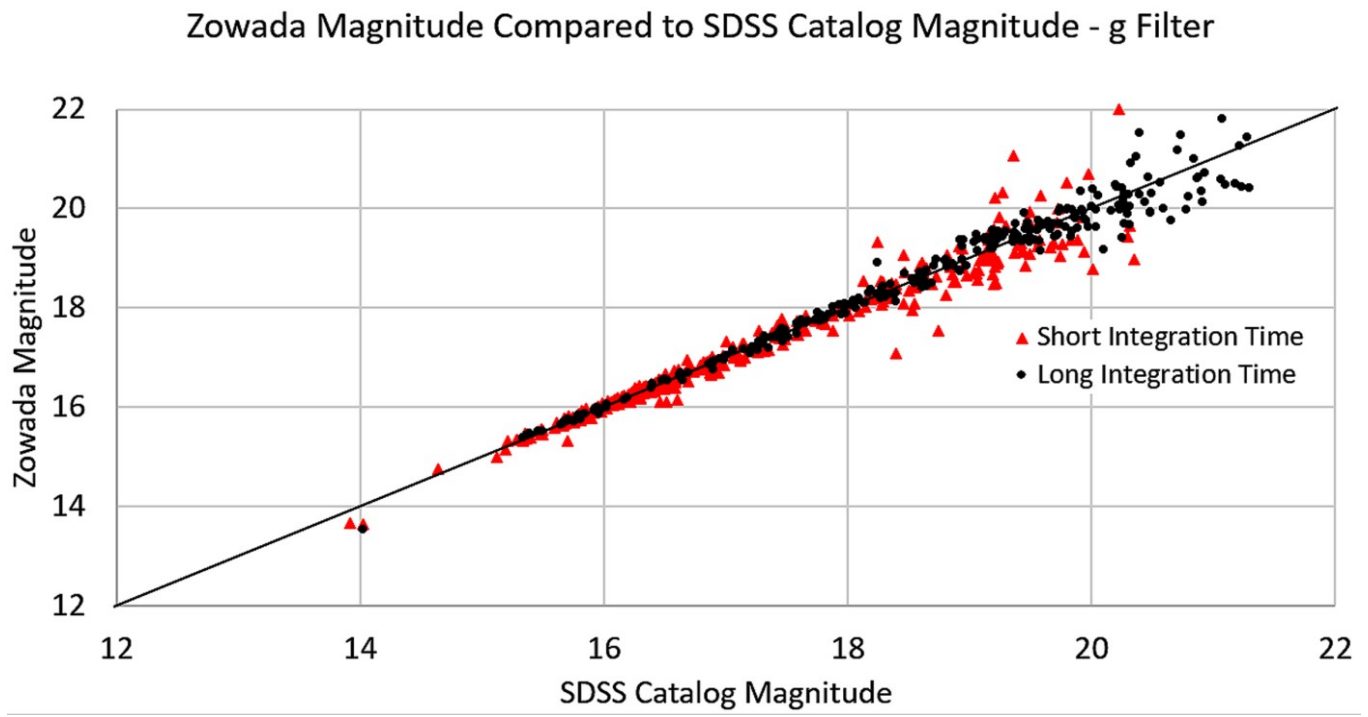

**Figure 7.** Magnitude measurements in g-band at Zowada compared to SDSS for 353 measured stars. Errors tend to increase with increasing magnitude and with i and z bands compared to r and g bands. Short integration times are generally 4 x 60 seconds or 4 x 90 seconds; long integration times are 16 x 90 seconds. SDSS photometry at ~14 magnitude is flagged as saturated.

*5.2 Limiting Magnitude for Zowada Observations*

Five-sigma limiting magnitude is evaluated based upon the noise level of science images. The magnitude is measured that corresponds to five times the noise level of the image background. For individual 90 second exposures, the 5-sigma detection limit is 17.6 magnitude in g, 16.7 magnitude in r, 16.5 in i and 15.2 in z band; for 16 x 90 second exposures these are 19.0, 18.1, 17.9 and 16.6 magnitude in griz bands, respectively.

Magnitudes may be measured with useful accuracy beyond the 5-sigma limiting magnitude. We assess a practical limit to magnitude measurements to be reached at that magnitude corresponding to the peak count of objects setting the best fit line. Beyond this magnitude, few additional stars may be resolved to build the regression line and the number of objects setting the best fit line begins to decrease. This is similar to one of the methods employed by the Canadian Astronomy Data Center for assessment of limiting magnitude in their MegaPipe image stacking pipeline (Canadian Astronomy Data Centre 2020). For Zowada, this practical limit is consistent with the 3-sigma limiting value magnitude. We find that the 3-sigma limit to observations for individual 90 second exposures is 18.2 magnitude in g, 17.2 magnitude in r, 17.1 magnitude in i and 15.7 in z-band. The 3-sigma limit to observations of longer integrations times of 16 x 90 second exposures are 19.5, 18.6, 18.4 and 17.1 magnitude in griz bands, respectively. Measurements to higher magnitudes are possible with increased uncertainties.

**6. Discussion**

We have shown that the Zowada Observatory provides accurate measurements at relatively short exposures. Accuracy compared to SDSS measurements improves for longer integration times. For integration of 16 x 90 second exposures, the five sigma detection limit of Zowada Observatory is 19.0 magnitude in g, 18.1 magnitude in r, 17.9 magnitude in i, and 16.6 magnitude in z. For a three sigma detection limit, measurements may be performed with greater uncertainties as deep as 19.5, 18.6. 18.4 and 17.1 magnitude in griz bands, respectively.

Zowada's capacity to perform measurements in multiple optical wavelengths expands the data available to Wayne State University's research and partnerships. Measurements and light curves in griz wavelengths provides valuable optical wavelength data to complement x-ray and radio and other observations. The Zowada Observatory can also provide measurements for targets of opportunity with daily or near daily cadence.

This work is funded by the Michigan Space Grant Consortium, Grant Number 80NSSC20M0124, Wayne State University, and the generous support of Russell Carroll.